\documentclass[sigconf]{acmart}

\usepackage{booktabs} % For formal tables

\usepackage{pifont}% http://ctan.org/pkg/pifont
\usepackage{xcolor}
\usepackage{listings}
\usepackage{todonotes}
\usepackage{hyperref}
\definecolor{codegreen}{rgb}{0,0.6,0}
\definecolor{codegray}{rgb}{0.5,0.5,0.5}
\definecolor{codepurple}{rgb}{0.58,0,0.82}
\definecolor{backcolour}{rgb}{0.95,0.95,0.92}
\lstdefinestyle{mystyle}{
    backgroundcolor=\color{backcolour},   
    commentstyle=\color{codegreen},
    keywordstyle=\color{magenta},
    numberstyle=\tiny\color{codegray},
    stringstyle=\color{codepurple},
    basicstyle=\ttfamily\footnotesize,
    breakatwhitespace=false,         
    breaklines=true,                 
    captionpos=b,                    
    keepspaces=true,                 
    numbers=left,                    
    numbersep=5pt,                  
    showspaces=false,                
    showstringspaces=false,
    showtabs=false,                  
    tabsize=2
}
\usepackage[T1]{fontenc}

\usepackage{colortbl}

\usepackage{hyperref}
\usepackage{threeparttable}

\usepackage{graphicx}
\usepackage{multirow}

\usepackage{caption}
\usepackage{subcaption}

\usepackage[tworuled, lined, linesnumbered]{algorithm2e}
\usepackage{tikz}
\usetikzlibrary{arrows, automata}
\usepackage{pgfplots}
\pgfplotsset{compat=newest}
\usepgfplotslibrary{fillbetween}

\usepackage{ifthen}
\definecolor{shadecolor}{rgb}{0.90,0.90,1}
\usepackage{tcolorbox}
\tcbset{colback=shadecolor, colframe=shadecolor, outer arc=3pt}

\begin{document}

\fancyhead{}

%\title{Do We Really Need All This? An Exploratory Study of JMH Parameterization in Open-Source Projects}
% \title{Performance Prediction by flow augmented AST and Graph Neural Networks}
\title{TEP-GNN: Accurate Execution Time Prediction of Functional Tests using Graph Neural Networks}

 \author{Hazem Peter Samoaa}
 \affiliation{%
     % Software Engineering Division
   \institution{Chalmers  University of Technology}
   %\streetaddress{}
   \city{Gothenburg}
   \country{Sweden}
   %\postcode{}
 }
 \email{samoaa@chalmers.se}

  \author{Antonio Longa}
  \affiliation{%
 %     % Software Engineering Division
   \institution{Fondazione Bruno Kessler and University of Trento}
% %   %\streetaddress{}
   \city{Trento}
   \country{Italy}
 %   %\postcode{}
  }
  \email{alonga@fbk.eu}

  \author{Mazen Mohamad}
  \affiliation{%
 %     % Software Engineering Division
   \institution{University of Gothenburg}
 %   %\streetaddress{}
   \city{Gothenburg}
   \country{Sweden}
 %   %\postcode{}
  }
  \email{mazen.mohamad@gu.se}
 
  \author{Morteza Haghir Chehreghani}
 \affiliation{%
     % Software Engineering Division
   \institution{Chalmers University of Technology}
   %\streetaddress{}
   \city{Gothenburg}
   \country{Sweden}
   %\postcode{}
 }
 \email{morteza.chehreghani@chalmers.se}

 \author{Philipp Leitner}
 \affiliation{%
     % Software Engineering Division
   \institution{Chalmers $|$ University of Gothenburg}
   %\streetaddress{}
   \city{Gothenburg}
   \country{Sweden}
   %\postcode{}
 }
 \email{philipp.leitner@chalmers.se}

% The default list of authors is too long for headers}
% \renewcommand{\shortauthors}{Laaber and Leitner}

\newcommand{\antonio}[1]{\textcolor{red}{ Antonio says: #1 }}

\newcommand{\approach}{~TEP-GNN\xspace}

%
% The code below should be generated by the tool at
% http://dl.acm.org/ccs.cfm
% Please copy and paste the code instead of the example below.
%
%
% \ccsdesc[500]{Software and its engineering~Software performance}
%
% % We no longer use \terms command
% %\terms{Theory}
%
% \keywords{Performance, Performance Testing, Information Mining}

\settopmatter{printacmref=false}

\begin{abstract}
Predicting the performance of production code prior to actually executing or benchmarking it is known to be highly challenging. In this paper, we propose a predictive model, dubbed \approach, which demonstrates that high-accuracy performance prediction is possible for the special case of predicting unit test execution times. \approach uses FA-ASTs, or flow-augmented ASTs, as a graph-based code representation approach, and predicts test execution times using a powerful graph neural network (GNN) deep learning model. We evaluate \approach using four real-life Java open source programs, based on 922 test files mined from the projects' public repositories. We find that our approach achieves a high Pearson correlation of  0.789, considerable outperforming a baseline deep learning model. However, we also find that more work is needed for trained models to generalize to unseen projects. Our work demonstrates that FA-ASTs and GNNs are a feasible approach for predicting absolute performance values, and serves as an important intermediary step towards being able to predict the performance of arbitrary code prior to execution.

\end{abstract}

\maketitle

 \section{Introduction}
\label{sec:introduction}
% we will write this section following the ‘CARS’: Creating a Research Space model
% \textbf{Move I :Establishing a territory\\
% a. showing that the general research area is important,central, interesting, etc.}\\

Performance is a critical quality property of many real-live software systems. Hence, performance modeling and analysis have gradually become an increasingly important part of the software development life-cycle. Unfortunately, predicting the performance of real-life production code is well-known to be a difficult problem -- predicting the absolute execution time of applications based on code structure is challenging  as it is a function of many factors, including the underlying architecture, the input parameters, and the application’s interactions with the operating system~\cite{Ramadan2021}. Consequently, works that attempted to predict absolute performance counters (e.g., execution time) for arbitrary applications from source code generally report poor accuracy~\cite{Meng2017,Narayanan2010}.

However, recent research has shown that predicting performance characteristics is indeed possible in more specialized contexts, via the application of modern machine learning architectures. For example, Guo et al. successfully predict the execution time of a specific untested configuration of a configurable system~ \cite{guo:13,guo:18}, Samoaa and Leitner have shown that the execution time of a benchmark with specific workload configuration can be predicted~\cite{samoaa:21}, and Laaber et al. have shown that a categorical classification of benchmarks into high- or low-variability is feasible~\cite{laaber:21}.

In this work, we demonstrate that another context where performance prediction is possible is the prediction of execution times of functional tests.
Test execution times are crucial in agile software development and continuous integration.
While individual test cases might have short execution times, software products often have thousands of test cases, which makes the total execution time in the build process high. Researchers have been working on solutions to speed up the testing process by optimizing the code or prioritizing test cases \cite{codeOptimization1,testPrio1,testPrio2,testPrio3}. The goal of this study is to provide the developers with predictions of the execution times of their test cases, and consequently giving them an early indication of the time required to run the cases in the build process. We believe that this would support decisions regarding code optimization and test case selection in early stages of the software life-cycle. 

In this paper, we propose an approach dubbed \approach (Test Execution Time Prediction using Graph Neural Networks) that makes use of structural features of test cases (the abstract syntax tree, AST). In \approach we enrich the AST with various types of edges representing data and control flow. Following Wang and Jin, we refer to this resulting graph as flow-augmented abstract syntax trees (FA-AST) ~\cite{Wang20}. We apply a graph neural network (GNN) deep learning model, more specifically GraphConv \cite{morris2019weisfeiler}, on the resulting FA-ASTs to predict execution times. We train and test our model on a dataset collected from four well-known open source projects hosted on GitHub: \textit{H2 database}\footnote{https://github.com/h2database/h2database}, a relational database,  \textit{RDF4J}\footnote{https://github.com/eclipse/rdf4j}, a project for handling RDF data, \textit{systemDS}\footnote{https://github.com/apache/systemds}, an Apache project to manage the data science life cycle, and finally the Apache remote procedure call library \textit{Dubbo}\footnote{https://github.com/apache/dubbo}. As labelled ground truth data, we collect 922 real test execution traces from these projects' publicly available build systems. 

\begin{figure*}[!ht]
    \centering
    \includegraphics[width=\textwidth]{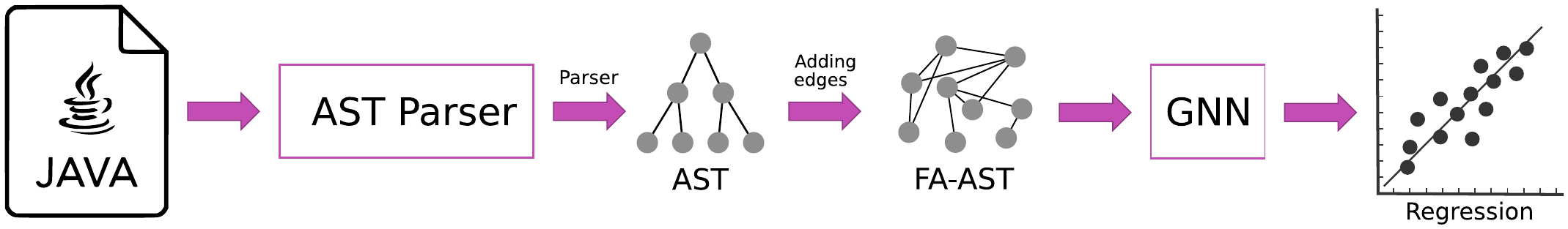}
    \caption{Schematic overview of the main phases of \approach. Java unit tests are parsed into ASTs, which get augmented with control and data flow edges. The resulting graph is then used as input for a GNN.}
    \label{fig:approach}
\end{figure*}

We conduct experiments with our \approach model to answer the following research questions: 

\begin{itemize}
    \item \textbf{RQ1:} How accurately can the absolute execution time of a test file consisting of one or multiple test cases be predicted using FA-ASTs and GNNs?
    \item \textbf{RQ2:} Does our usage of GraphConv improve execution time prediction compared to a baseline using Gated Graph Neural Networks (GGNN), as frequently used in previous software engineering research~\cite{Fernandes2018, Allamanis2017}?  
    \item \textbf{RQ3:} How well are models trained on a subset of projects transferable to unseen other projects? Can a \approach model trained on a subset of projects predict the execution time of tests in a project that was not used during training?
\end{itemize}

Our results show that using \approach, test execution time can be predicted with a very high 
prediction accuracy (Pearson correlation of 0.789). Further, we show that our usage of GraphConv indeed improves the model significantly over GGNN. Finally, we show that trained models are \emph{not} easily transferable. We conclude that test execution times can indeed be predicted using GNN models with high accuracy, even based on performance counters that have been collected "in the wild" by real projects (as opposed to performance measurements collected on a dedicated performance testing machine). Even though test cases are shorter and structurally simpler than arbitrary programs, we see our results as an important stepping stone towards the prediction of the performance of arbitrary software systems prior to execution. However, more work is needed to train general models that can be applied to arbitrary, unseen projects.

 \section{The TEP-GNN Approach}
\label{sec:approach}

In this section, we introduce \approach. We first provide a general overview of the model and discuss the problem addressed in this paper, followed by a detailed discussion of the main components of \approach (FA-ASTs and the machine learning pipeline based on the GraphConv~\cite{morris2019weisfeiler} higher order GNN).

\subsection{Approach Overview}
Our goal in this paper is to predict the execution time of test cases based on static code information alone, i.e., without access to prior benchmarking of the test case or dynamic analysis data. The general procedure of our \approach approach is sketched in  Figure~\ref{fig:approach}. To process a test file, we first parse it into its AST. Next, we build a graph representation (FA-AST) by adding edges representing control and data flow to the AST. We then initialize the embeddings of FA-AST nodes and edges before jointly feeding a vectorized FA-AST into a GNN. We use a 3-layer higher order graph convolution neural network to predict the execution time. Each layer is followed by a ReLU activation function. Since GNN learns node embedding, we use global max pooling to compute a graph embedding. Finally, the graph embedding goes into two Linear layers with a ReLU and a sigmoid activation function to perform the prediction of the test execution time. To train our model we use the mean square error loss.

% We now briefly formally introduce the problem that \approach solves, followed by in-depth discussion of the two main parts of the approach - building FA-ASTs and using GNNs to predict test execution time.

% first introduce an overview of our proposed approach based on program graphs and graph neural networks. Then we define the problem. Next, we describe the process of building a graph representation: flow-augmented abstract syntax tree (FA-AST) for code fragments. We then explain the technical details of our neural network models: Higher order graph neural networks . 

\subsection{Problem Definition}
Given a test file (source code containing one or multiple test cases) $C_i$ and the corresponding run-time value $X_i$ (execution time of all test cases in the file), for a set of test files with known execution times
 we can build a training set $D = {(Ci, Xi )}$. We aim to train a deep learning model for learning a function $\phi$ that maps a test file $C_i$ to a feature vector $v$  mapped to the corresponding value $X_i$.

\subsection{Building Flow-Augmented Abstract Syntax Trees}
\label{sec:fa-ast}
Recent studies \cite{samoaa2022} emphasize the importance of the code representation when using deep learning in software engineering. Hence, and
given the complexity of predicting performance, prediction based on the syntactical information extracted from ASTs alone is not sufficient to achieve high-quality predictions. In \approach, the basic structural information provided by the AST is enriched with semantic information representing  data and control flow. Consequently, the tree structure of the AST is generalized to a (substantially richer) graph, encoding more information than code structure alone. This idea is based on the earlier work by Wang and Jin~\cite{Wang20}, who have also introduced the term FA-AST for this kind of source code representation.

\subsubsection{AST Parsing}

\begin{lstlisting}[float=h!, language=Java, caption=A Simple JUnit 5 Test Case, label=java:example]
package org.myorg.weather.tests;

import static
    org.junit.jupiter.api.Assertions.assertEquals;
import org.myorg.weather.WeatherAPI;
import org.myorg.weather.Flags;

public class WeatherAPITest {
    
    WeatherAPI api = new WeatherAPI();
    
    @Test
    public void testTemperatureOutput() {
        double currentTemp = api.currentTemp();
        Flags f = api.getFreezeFlag();
        if(currentTemp <= 3.0d)
            assertEquals(Flags.FREEZE, f);
        else
            assertEquals(Flags.THAW, f);
    }
}
\end{lstlisting}

We demonstrate our approach for constructing FA-ASTs for test files using the example of a Java JUnit 5 test case (see Listing~\ref{java:example}). In this  example, a single test case \texttt{testTemperature\-Output()} is presented that tests a feature of an (imaginary) API. As common for test cases, the example is short and structurally relatively simple. Much of the body of the test case consists of invocations to the system-under-test and calls of JUnit standard methods, such as \texttt{assertEquals}. We speculate that these properties make predicting test execution time a more tractable problem than predicting performance of general-purpose programs, which previous authors have argued to be extremely challenging~\cite{Meng2017,Narayanan2010}.

\begin{figure}[!h]
    \centering
    \includegraphics[width=\columnwidth]{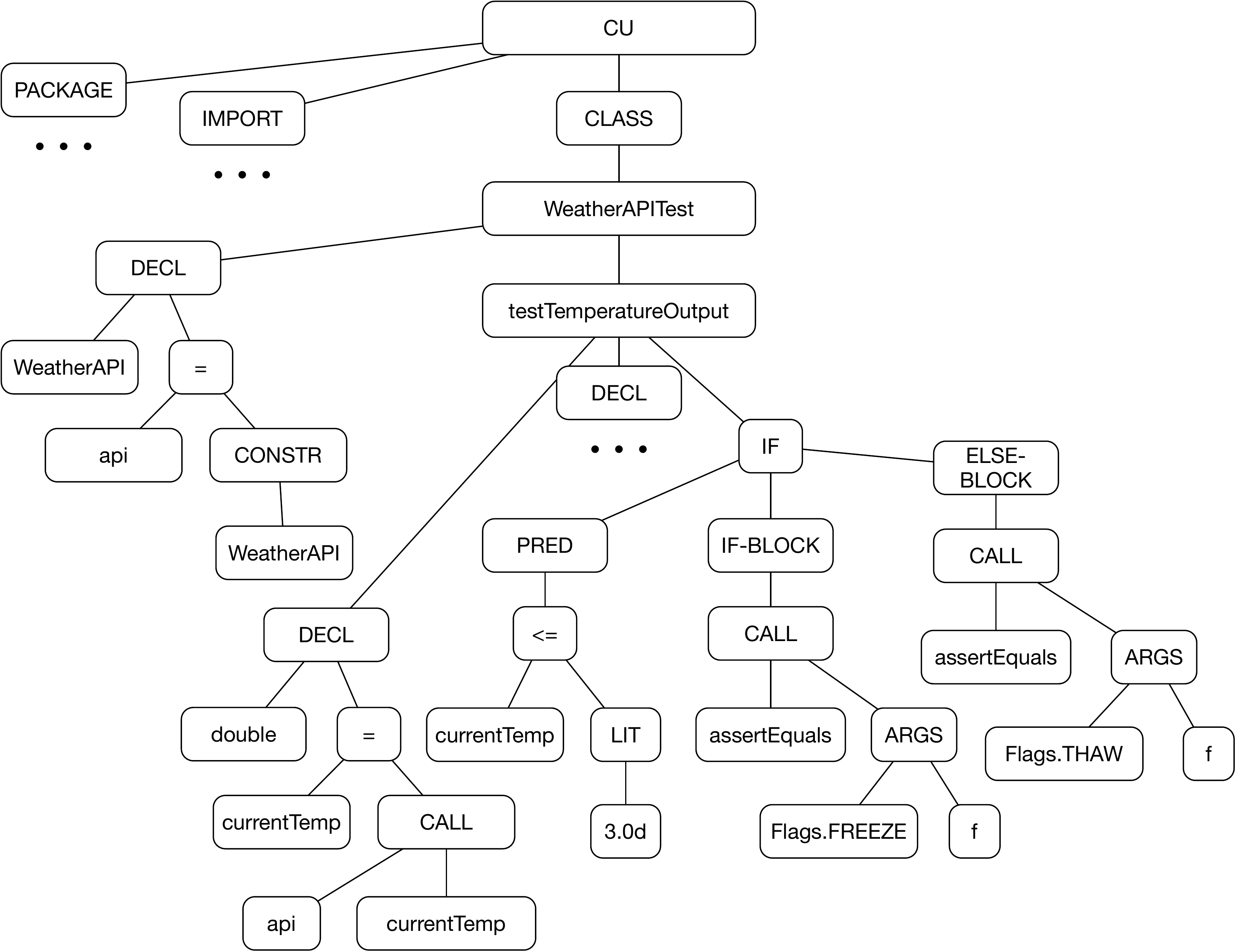}
    \caption{Simplified abstract syntax tree (AST) representing the illustrative example presented in Listing~\ref{java:example}. Package declarations, import statements, as well as the declaration in Line 15 are skipped for brevity.}
    \label{fig:AST}
\end{figure}

A (slightly simplified) AST for this illustrative example is depicted in Figure~\ref{fig:AST}. The produced AST does not contain purely syntactical elements, such as comments, brackets, or code location information. We make use of the pure Python Java parser javalang\footnote{https://pypi.org/project/javalang/} to parse each test file, and use the node types, values, and production rules in javalang to describe our ASTs.

\subsubsection{Capturing Ordering and Data Flow}

In the next step, we augment this AST with different types of additional edges representing data flow and node order in the AST. Specifically, we use the following additional flow augmentation edges, in addition to the \textbf{AST child} and \textbf{AST parent} edges that are produced readily by AST parsing:

\textbf{FA Next Token} (b):\\
This type of edge connects a terminal node (leaf) in the AST to the next terminal node. Terminal nodes are nodes without children. In Figure~\ref{fig:AST}, an FA Next Token edge would be added, for example, between \texttt{WeatherAPI} and {api}.

\textbf{FA Next Sibling} (c):\\
This connects each node (both terminal and non-terminal) to its next sibling. This allows us to model the order of instructions in an otherwise unordered graph structure. In Figure~\ref{fig:AST}, such an edge would be added, for example, connecting the first usage of \texttt{api} and with the \texttt{CONSTR} node (representing a Java constructor call).

\textbf{FA Next Use} (d):\\
This type of edge connects a node representing a variable to the place where this variable is next used. For example, the variable \texttt{api} is declared in Line 10 in Listing~\ref{java:example}, and then used next in Line 14. 

\begin{figure*}
    \centering
    \includegraphics[width=\textwidth]{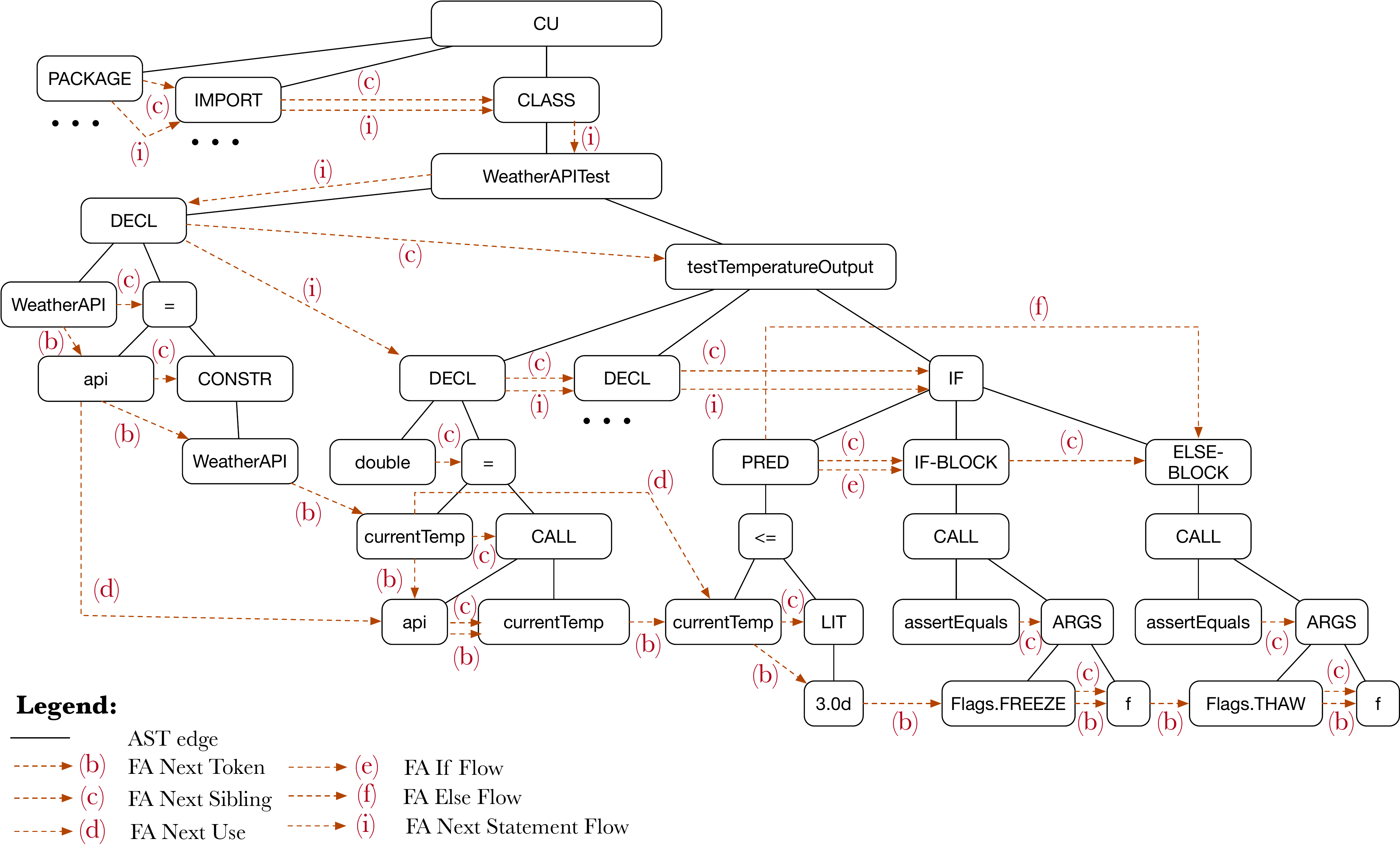}
    \caption{Flow-Augmented AST (FA-AST) for the example presented in Listing~\ref{java:example}. Solid lines represent AST parent and child edges, and dashed lines different types of flow augmentations.}
    \label{fig:FAAST}
\end{figure*}

Figure \ref{fig:FAAST} shows an example augmenting the AST in Figure \ref{fig:AST} (and, consequently, the example test case in Listing~\ref{java:example}). Solid black lines indicate the AST parent and child relationships (for simplicity indicated through a single arrow, read from top to bottom). Red dashed arrows refer to the new edges added to represent the data and control flow in the FA-AST, with letter codes indicating the edge type. Terminal nodes are connected with FA Next Token edges (b), modelling the order of terminals in the test case. Similarly, the ordering of siblings is modelled using FA Next Sibling edges (c). Finally, data flow is modelled by connecting each variable to their next usage via FA Next Use edges (d). Edge types (e), (f), and (i) represent a control flow statement, which will be discussed in the following. Multiple edges of different types are possible between the same nodes. For example, the terminal nodes \texttt{Flags.FREEZE} and \texttt{f} are connected via both, an FA Next Token (b) and an FA Next Sibling (c) edge.

\subsubsection{Capturing Control Flow}
In a second augmentation step, we now add further edges representing the control flow in the test cases. We currently support \emph{if} statements, \emph{while} and \emph{for} loops, as well as \emph{sequential execution}. We currently do not support \emph{switch} statements or \emph{do-while} loops, as these are less common in test cases. Test files containing these elements will still be parsed successfully, but these control flow constructs will not be captured by the FA-AST. Specifically, the following further edges are added:
% An overview over the additional edges introduced by these control flow statements is given in Figure~\ref{fig:CFG}.  

% \begin{figure}[h!]
%     \centering
%     \includegraphics[width=\columnwidth]{figures/flow_augmentation.pdf}
%     \caption{Additional flow augmentation edges introduced to capture control flow, namely if statements, while loops, for loops, and sequential execution.}
%     \label{fig:CFG}
% \end{figure}

\textbf{FA If Flow} (e):\\
This type of edge connects the predicate (condition) of the if-statement with the code block that is executed if the condition evaluates to \texttt{true}. Every if statement contains exactly one such edge by construction.

\begin{figure}[!h]
    \centering
    \includegraphics[width=0.75\columnwidth]{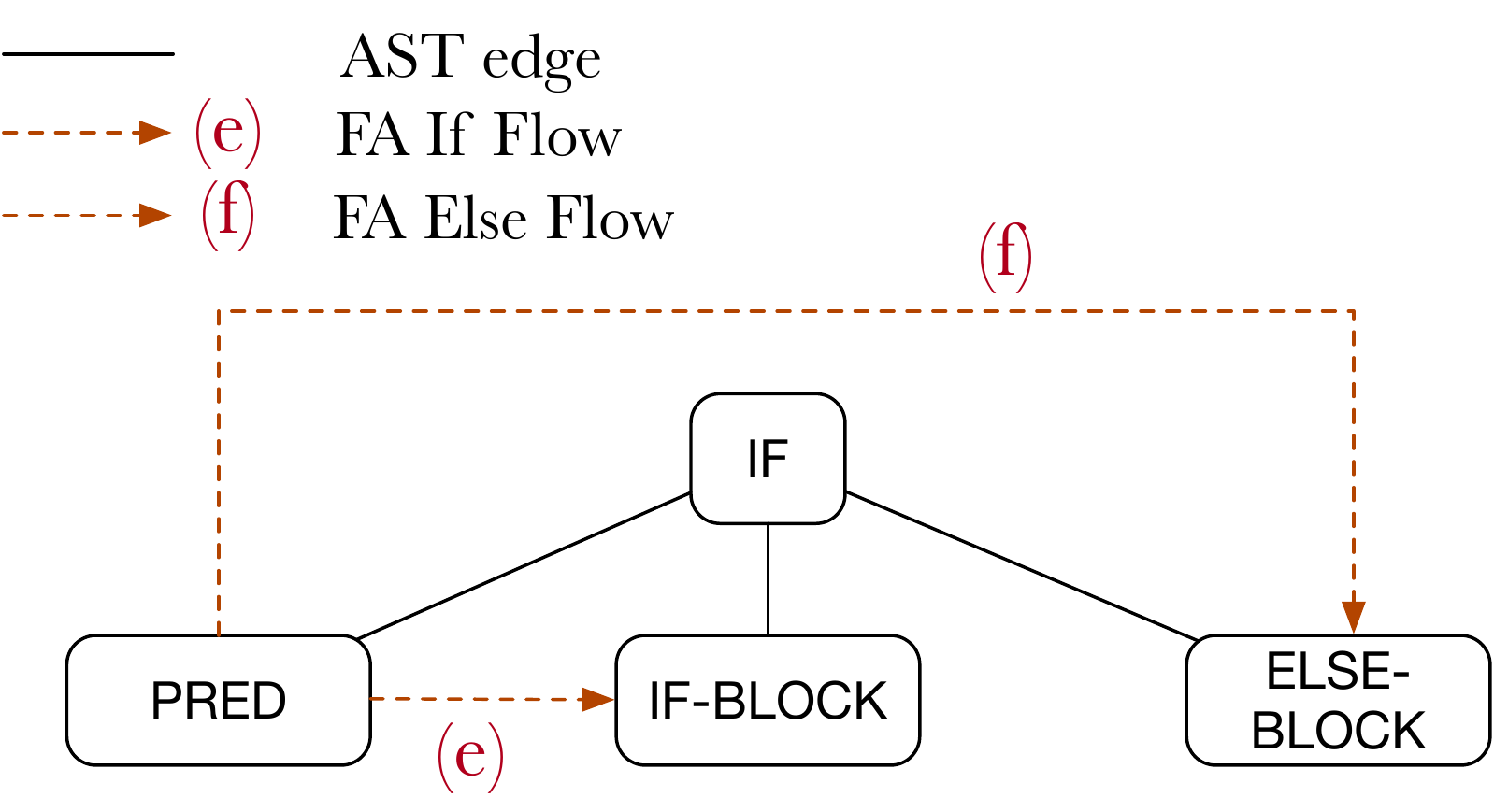}
    \caption{Capturing the control flow of if statements.}
    \label{fig:if}
\end{figure}

\textbf{FA Else Flow} (f):\\
Conversely, this edge type connects the predicate to the (optional) else code block. Figure~\ref{fig:if} depicts how if-statements are modelled.

\begin{figure}[!h]
    \centering
    \includegraphics[width=0.75\columnwidth]{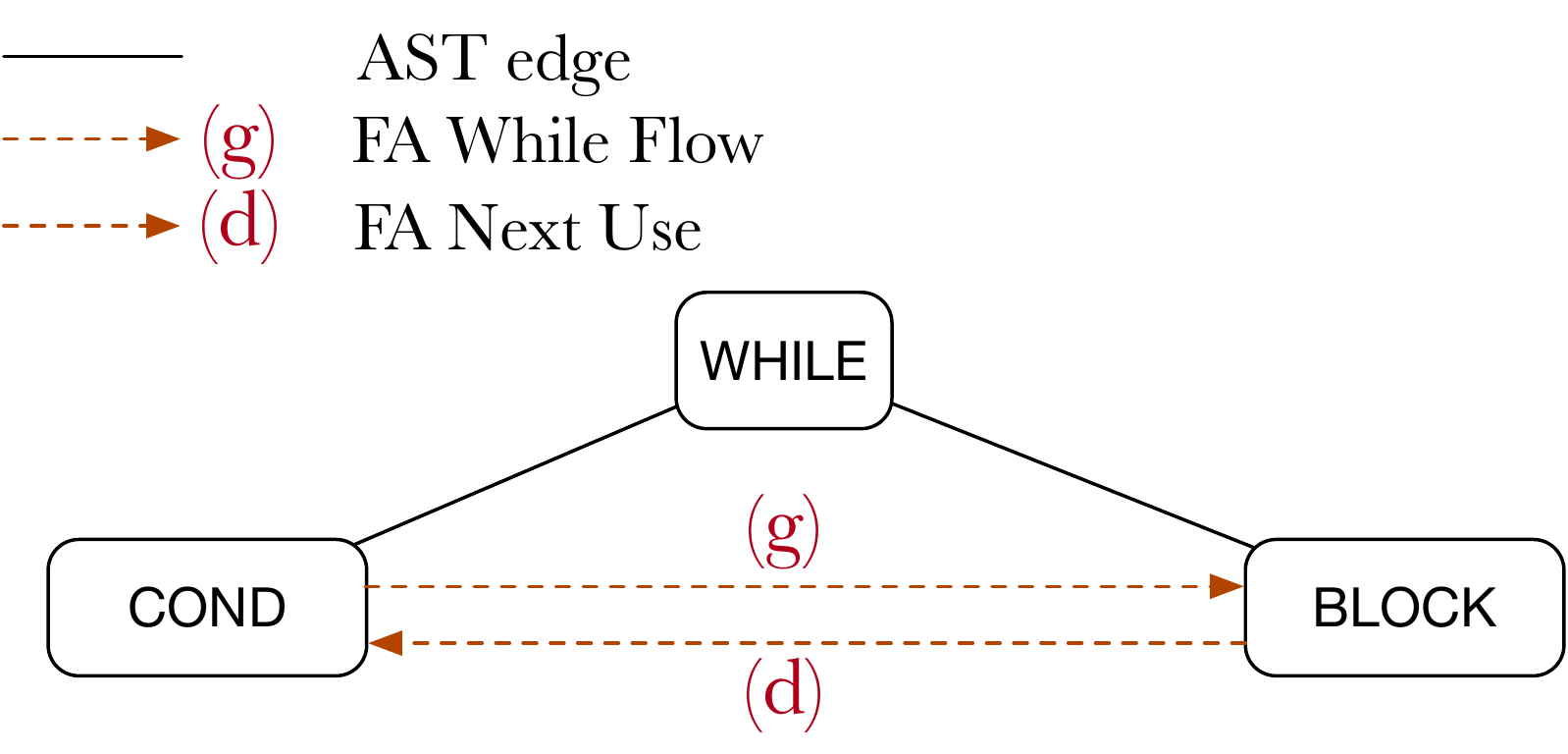}
    \caption{Capturing the control flow of while loops.}
    \label{fig:while}
\end{figure}

\textbf{FA While Flow} (g):\\
A while loop essentially entails two elements - a condition and a code block that is executed as long as the condition remains \texttt{true}. We capture this through a FA While Flow (g) edge connecting the condition to the code block, and an FA Next Use (d) edge in the reverse direction. The latter is used to model the next usage of a loop counter. Figure~\ref{fig:while} shows this.

\begin{figure}[!t]
    \centering
    \includegraphics[width=0.75\columnwidth]{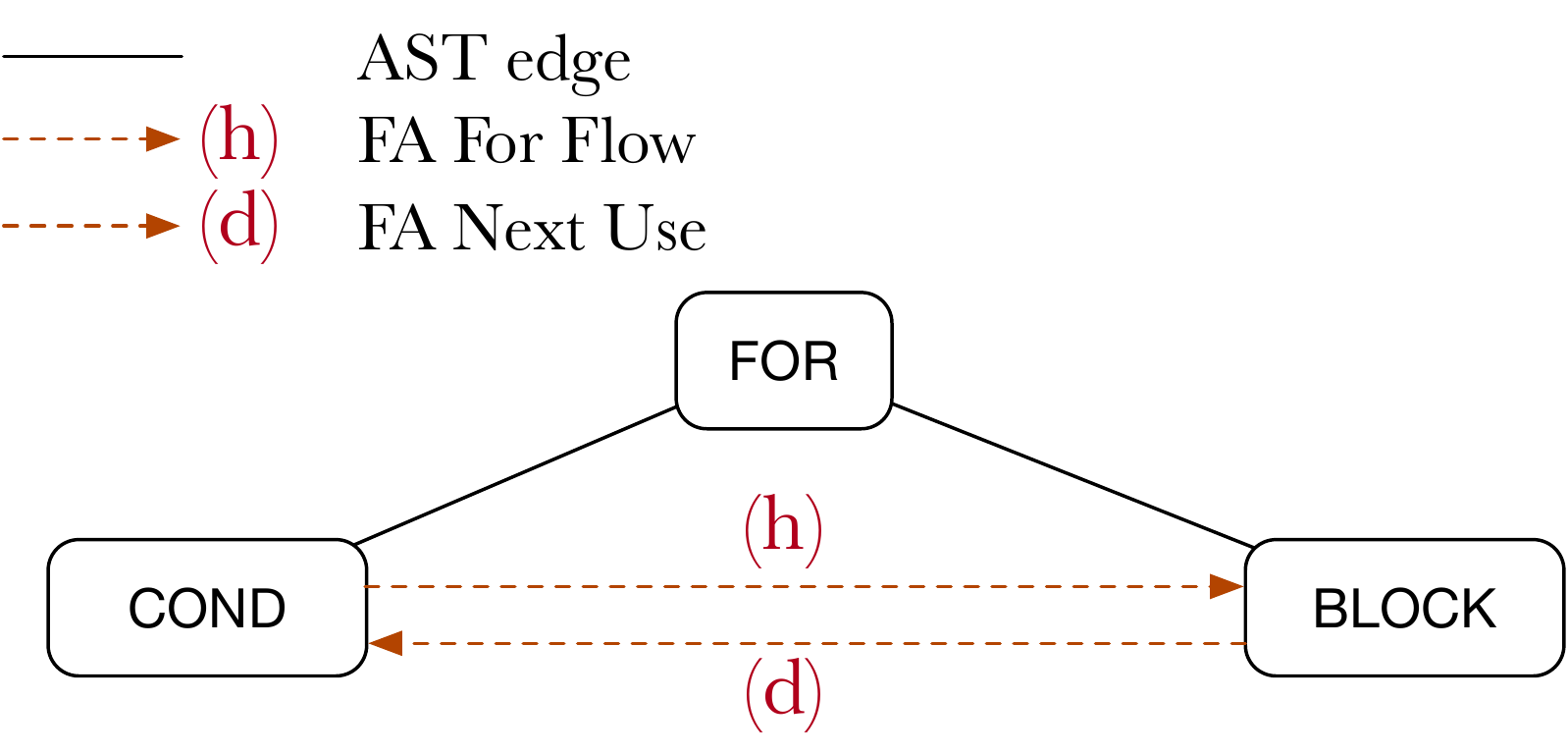}
    \caption{Capturing the control flow of for loops.}
    \label{fig:for}
\end{figure}

\textbf{FA For Flow} (h):\\
For loops are conceptually similar to while loops. We use FA For Flow (h) edges to connect the condition to the code block, and an FA Next Use (d) edge in the reverse direction. Similar to the modelling of while-loops, FA Next Use (d) relates to the usage (typically incrementing) of a loop counter. This is again illustrated in Figure~\ref{fig:for}.

\begin{figure}[h!]
    \centering
    \includegraphics[width=0.75\columnwidth]{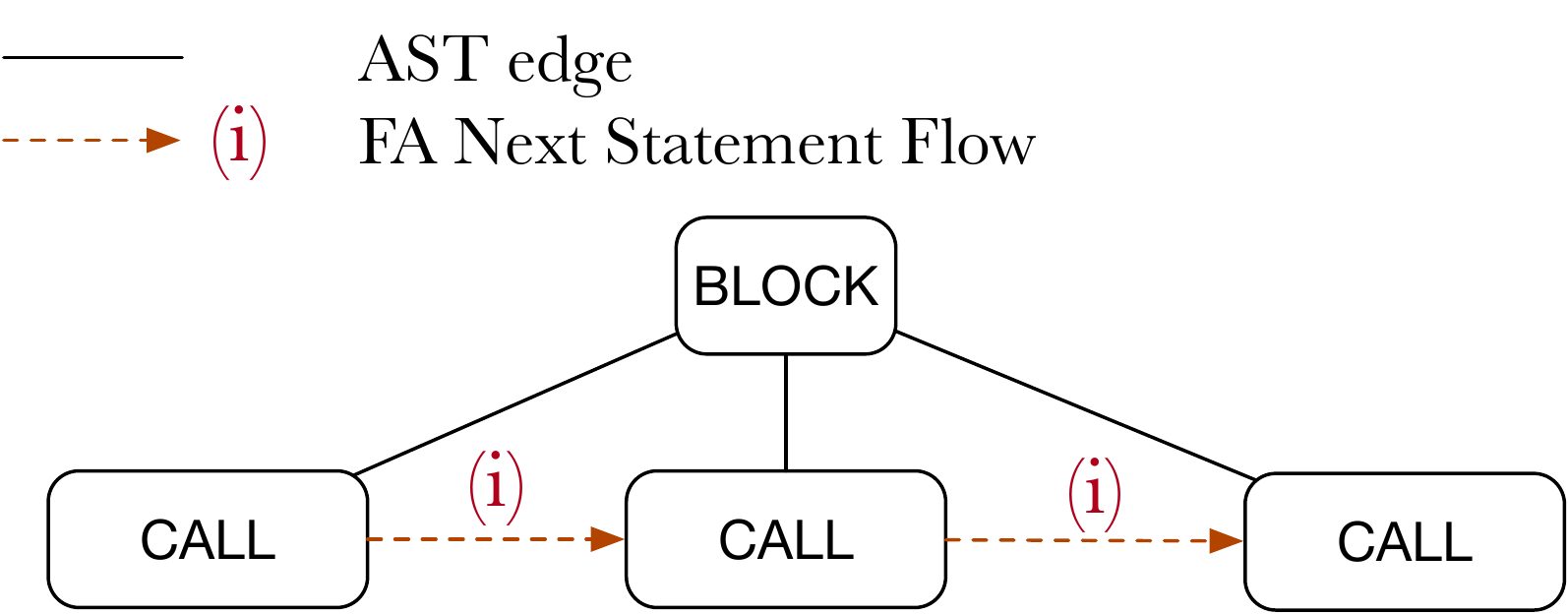}
    \caption{Capturing the control flow of sequential execution.}
    \label{fig:seq}
\end{figure}

\textbf{FA Next Statement Flow} (i):\\
In addition to the control flow constructs discussed so far, Java of course also supports the simple sequential execution of multiple statements in a sequence within a code block. FA Next Statement Flow edges (i) are used to represent this case. Different from the constructs discussed so far, a code block can contain an arbitrary number of children, and the FA Next Statement Flow edge is always used to connect each statement to the one directly following it. An illustration is shown in Figure~\ref{fig:seq}.

Referring back to Figure \ref{fig:FAAST}, two types of control flow annotations are visible - the modelling of the if-statement in lines 16 to 19 of the test case on the right-hand side, and various sequential executions (FA Next Statement flow (i) ) edges. Further note how flow annotation adds a large number of edges to even a very small AST, transforming the syntax tree into a densely connected graph. This rich additional information can be used in the next step by our GNN model to predict highly accurate test execution times.

% The rest are for the control statement in our example, which is \textit{If statement}. So, we added two edges which are (e) that represent \textit{If Flow} when the if condition is true, and (f) represents \textit{Else Flow} when the first if condition is not true.

\begin{figure*}
    \centering
    \includegraphics[scale = 0.8]{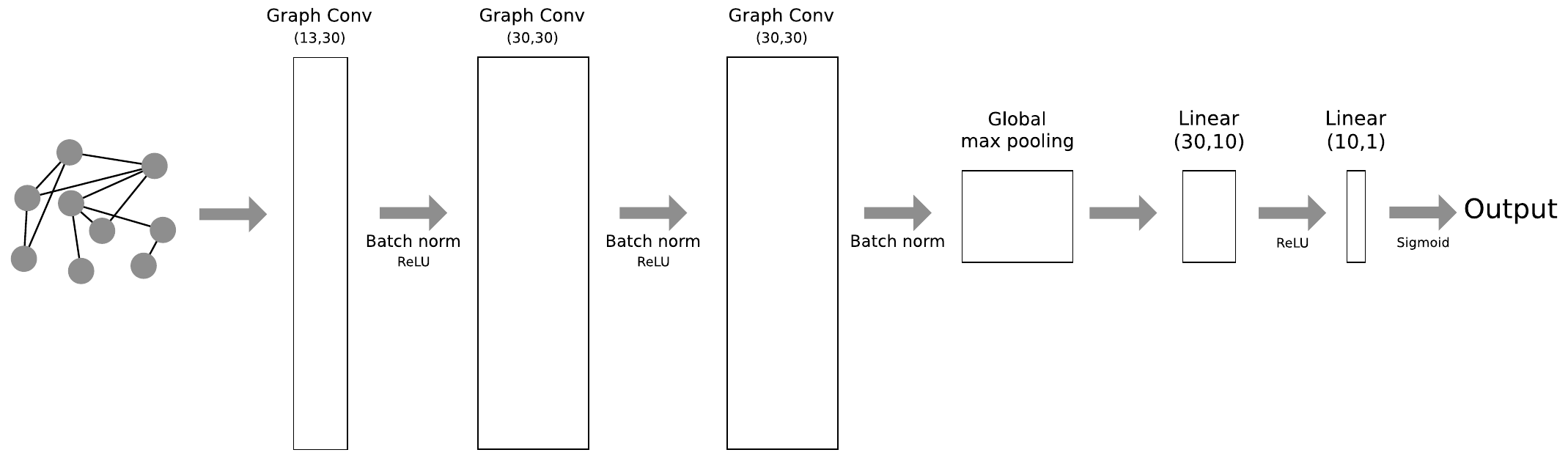}
    \caption{Architecture of the GNN Model used in \approach.}
    \label{fig:GNN}
\end{figure*}

\subsection{GNN Model for Test Execution Time Prediction}
% \todo[inline]{Philipp pleaase read this section and check if you understand all the details}
%which can take higher-order 
Once the FA-AST graph has been built for a test file using the three steps discussed above, we use a higher order GNN model to predict the execution time of the Java code.

Graphs are complex structures, verifying if two graphs are identical (also known as isomorphism test) is an important and difficult task. It is unknown if the problem can be solved in polynomial time or it is computationally intractable for large graphs. 
%Verifying if two graphs are identical (also known as isomorphism test) is an important and difficult task. It is unknown if the problem can be solved in polynomial time nor to be NP-complete. 
A fast heuristic to verify if two graph are the same is the $k$-Weisfeiler-Leman test \cite{weisfeiler1968reduction}. The algorithm produces for each graph a canonical form. Then, if the canonical forms of two graphs are not equivalent,  the graphs are not considered isomorphic. However, there is the possibility that two non-isomorphic graphs share a canonical form. Thereby, this test might not provide a conclusive evidence that two graphs are isomorphic. 
Morris et al. \cite{morris2019weisfeiler} prove that a GNN network can be as powerful as the $k$-Weisfeiler-Leman test with $k$ equal to 1, in which the canonical form propagates the information by nodes. With $k$ greater than 1, the information is propagated among substructures of order $k$. Morris et al. \cite{morris2019weisfeiler} also propose a higher order graph convolution layer ($k$-GNN), wherein messages are exchanged among nodes, edges and substructure with tree nodes (triads). Once messages are exchanged among substructures, each node has a latent representation. In order to predict property of the graphs (i.e., the execution time of a graph representing Java code), node embeddings are globally aggregated (pooling step) with an ordering invariant function (i.e. sum, max, mean). In particular, $k$-GNN is defined as: given is an integer $k$ the $k$-element subset $[V(G)]^k$ over $V(G)$. Let $s=\{s_1,s_2,\dots \}$ be $k$-set in $[V(G)]^k$, then the \textit{neighborhood} of $s$ is defined as:
\begin{equation}\label{eq:neigh}
    N(s) = \{t \in [V(G)]^k | |s\cup t |=k-1\}
\end{equation}
In Equation \ref{eq:neigh}, the neighbour of a $k$-set is defined as the set of $k$-set such that the intersection of their cardinality is equal to $k-1$.
The local neighborhood is defined as:
\begin{equation}\label{eq:loc_nei}
    N_L(s) = \{t\in N(s) | (u,w) \in E(G) \text{ with } u \in s/t \text{ and } w \in t/s\}
\end{equation}
The local neighborhood defined in Equation \ref{eq:loc_nei} is a subset of the neighbour (eq. \ref{eq:neigh}). Finally, the $k$-GNN is defined as:
\begin{equation}
    f_{k,L}^{(l)}(s) = \sigma(f_{k,L}^{(l-1)}(s)\cdot W_1^{(t)} + \sum_{u\in N_{L}(s)} f_{k,L}^{(t-1)}(u)\cdot W_2^{(t)} )
\end{equation}
The $l$-th layer of the $k$-GNN computes an embedding of $s$, using the non linear activation function $\sigma$ of the summation over the substructure itself in the previous layer (i.e., layer $l-1$) and the summation over the previous layer embedding of each local neighbor of the substructure $s$.
A scheme of the deep learning architecture we use in \approach is shown in Figure~\ref{fig:GNN}. Due to the proven superiority of higher order GNN with respect to classic GNN~\cite{morris2019weisfeiler}, we adopt \textit{GraphConv}, a PyTorch-geometric implementation of the higher order Graph Neural Network layer. The GNN is composed of three convolutional layers, each layer has a \textit{GraphConv} layer with an \textit{ReLU} activation function and a \textit{batch normalization}. Then node embeddings are converted into a graph embedding using a \textit{global max pooling} layer. Finally, the predicted execution time is predicted using two multi-layers perceptrons with \textit{ReLU} and \textit{sigmoid} activation functions. We use the Mean Square Error as the loss function. In order to make the learning easier for the neural network, and to reduce the optimization problems, real execution times have been normalized between 0 and 1.\\
% \todo[inline]{Antonio please add some mathematical equations related to the high order \textit{GraphConv} since many papers that invest GNN in SE do that. Please look at \cite{Wang20} Page 5 first column}
 \section{Evaluation}
\label{sec:eval}

\begin{table*}[tb]
\centering
\caption{Overview of study subjects.}
\begin{tabular}{@{}p{2cm}p{3.8cm}p{1.5cm}p{1.5cm}p{2.5cm}p{2.5cm}@{}}
 \toprule
 \textbf{Project} & \textbf{Description}& \textbf{Test files} & \textbf{Test runs} & \textbf{Number of nodes} & \textbf{Vocabulary size} \\
 
 \midrule
 systemDS & \small{Apache Machine Learning system for data science lifecycle} & 127 & 1321 & 110651 & 3161 \\
 \rowcolor[rgb]{0.953,0.953,0.953} H2 & \small{Java SQL database} & 194 & 1391& 405706 & 17972\\
 Dubbo & \small{Apache Remote Procedure Call framework} & 123 & 524 & 75787 & 4499 \\
 \rowcolor[rgb]{0.953,0.953,0.953} RDF4J & \small{Scalable RDF processing for Java} & 478 & 1055 & 214436 & 10755 \\

 \midrule
 \textbf{Total} & & \textbf{922} & \textbf{4291} &\textbf{806580} &\textbf{36387}  \\
 \bottomrule
\end{tabular}
\label{tab:dataProjects}
\end{table*}

We now present the results of an experimental evaluation of \approach based on open source Java projects. As training and test data we make use of existing test suite execution traces from the study subjects' build systems. A replication package containing the scripts used to implement the \approach approach, all data used in the evaluation, as well as all analysis scripts, are available~\cite{hazem_peter_2022_7003881}.

\subsection{Dataset} \label{dataset}
% Current program performance analysis tools can be categorized into static and dynamic. Dynamic (runtime) analysis is performed by executing the target program and measuring metrics of interest, e.g., time or hardware performance counters. By contrast, static analysis operates on the source or binary code without actually executing it. Thus, we relied on the dynamic analysis data retrieved from four executed projects in order to build a predictive model based on static analysis. 
 
%\todo[inline]{Mazen: Talk about the process of data collection}
Related studies in performance engineering frequently collect their own performance data, for example by repeated execution of the projects on a researcher's laptop~\cite{alcocer:16}, in cloud virtual machines~\cite{laaber:19}, or on controlled hardware~\cite{schulz:21}. To increase the realism of the study we have chosen a different approach -- we harvest existing execution traces from an open source build system (GitHub), and extract test execution times from this public data. This data represents actual, real-life test execution times. However, we do not have the option to collect more data on-demand, and we do not know what precise hardware has been used to collect the data.

% The dataset used in this study is extracted from GitHub repositories. 
%
To collect the data, we searched for projects to serve as study subjects. We applied the following selection criteria: \emph{(i)} projects written in Java; \emph{(ii)} available on GitHub; \emph{(iii)} include test results published on GitHub; and \emph{(iv)} use GitHub shared runners as build system.

Based on these criteria, we selected four projects of diverse application domains, i.e., databases, web servers, and data science life-cycle (systemDS, H2, Dubbo, and RDF4J). These are depicted in Table \ref{tab:dataProjects}. The first column shows the project's name, the second provides a brief description of the project. The third column shows the number of distinct test files extracted from the project. As for the fourth column, it shows the total number of runs performed in the testing job. The last two columns show the total number of tokens in the entire project test files and the vocabulary size (the number of distinct nodes in the graphs). We observe that RDF4J, a triplestore database used in semantic web contexts~\cite{samoaa-pipeline:2021}, contains more test files than the other projects. For the H2 relational database and systemDS we were able to collect the most test runs. Finally, it should be noted that H2 has the highest code density as measured by the number of nodes and the resulting vocabulary size by a wide margin. This indicates that H2 tests are generally larger and more complex than the test cases in the other study subjects.

All data was extracted from GitHub-hosted runners, which are virtual machines hosted by GitHub with the GitHub Actions runner application installed. All shared runners can be assumed to use the same hardware resources, which is available at GitHub's website\footnote{https://docs.github.com/en/actions/using-github-hosted-runners/about-github-hosted-runners\#supported-runners-and-hardware-resources} and each job runs in a fresh instance of the virtual machine. Additionally, all jobs from which the data is extracted uses the same operating system, specifically Ubuntu 18.04.
This allows us to minimize bias introduced by variations in execution environment or hardware. 

For collecting test execution traces we looked at the latest successful action workflow run for each project. We then extracted the run times from the test report in the workflow, and mined the corresponding source code files from the respective project repositories in order to feed them to the parser. 
For H2, some test cases are run several times during the same build job. In these cases, we recorded the average of the run times. As the execution times of tests can vary dramatically between and within projects, to increase the efficiency of the model training, we normalize each execution time to interval $[0;1]$. Hence, our final dataset includes distinct test files, each associated with one runtime value between 0 and 1. Then after model training, we denormalize the runtime value and present the results based on the original values. 

%
%Each source code file includes one or multiple test cases. 
%
Table \ref{tab:controlflow} indicates how prevalent different control flow nodes were in the test cases of our study subjects.
For all projects, block statements are the most frequent control
flow construct, since sequential executions widely exist in nearly all programs. For loops are substantially more common than while loops, and if statements are also frequent. Do-while loops and switch statements, which are currently unsupported by \approach, are both quite rare in the tests of our subjects (not shown in the table).

\begin{table}[h!]
\centering
\caption{Occurrences of Control Flow Nodes in Each Project }
\begin{tabular}{@{}p{2cm}p{1.2cm}p{1.7cm}p{1cm}p{1cm}@{}}
 \toprule
 \textbf{Control Flow Statement} & \textbf{systemDS}& \textbf{H2} & \textbf{Dubbo} & \textbf{RDF4J} \\
 
 \midrule
 If Statement & 166 & 1322 & 53 & 161 \\
 \rowcolor[rgb]{0.953,0.953,0.953} While Loop & 2 & 222 &3 & 22\\
 ForStatement & 196 & 1114 &42 &158\\
 \rowcolor[rgb]{0.953,0.953,0.953} Block Statement & 293 & 2900 & 116&395\\
% Do While Loop & 0 & 5 & 0&0\\
%  \rowcolor[rgb]{0.953,0.953,0.953} Switch Statement & 50 & 49 &0 &0\\
\midrule
 \textbf{Total} &\textbf{707} & \textbf{5612} & \textbf{214} & \textbf{736}   \\
 \bottomrule
\end{tabular}
\label{tab:controlflow}
\end{table}

\subsection{Results}
\label{sec:results}
In this section, we investigate the results of applying \approach to our dataset, answering RQ1 -- RQ3 introduced in Section~\ref{sec:introduction}.

\subsubsection{RQ1: Quality of Predictions}
In order to answer the first research question, we 
combine the collected data for all projects into one dataset entailing 922 code fragments and associated normalized execution times. After that, we apply \approach as discussed in Section~\ref{sec:approach}.
For model training, we split the dataset into train and test sets using $80\%$ and $20\%$, respectively. Each network is trained for 100 epochs. As optimizer we use Adam~\cite{Adam} with a learning rate $= 0.001$. \textcolor{black}{To evaluate the results of our model, we use a Pearson correlation metric, a measure of linear correlation between two sets of data. In addition, as a loss function, we use mean squared error, which is the average squared difference between the estimated and actual values.}  All experiments have been executed in a machine equipped with a GeForce 940MX graphics card and 16GB of RAM. 

\begin{table}[h!]
\centering
\caption{Results of the \approach on the entire dataset  }
\begin{tabular}{p{2cm}p{2cm}}
 \toprule
 \textbf{Pearson correlation} & \textbf{Mean squared error }\\
 
 \midrule
 0.789 & 0.017 \\
 \bottomrule
\end{tabular}
\label{tab:RQ1}
\end{table}

Results are shown in Table \ref{tab:RQ1}. Our model trained on FA-AST is able to predict test execution times with a very high accuracy, as can be seen in the Pearson correlation (between predicted and actual execution times in the test data set) of 0.789, and a mean squared error of 0.017.
These results substantially outperform the accuracy values reported in previous studies that attempted to predict absolute software performance counters~\cite{Meng2017,Narayanan2010}. We argue that the key innovation that enables this high accuracy is the combination of FA-AST as a powerful code representation model and GraphConv as a modern GNN.

\subsubsection{RQ2: Comparison of \approach Against a Baseline GNN}
To validate the suitability of our approach and the selected GNN model, we compare it to a commonly used GNN baseline, called Gated Graph Neural Networks (GGNN)\cite{Li2015}. GGNNs are widely used in studies that attempt to learn code semantics~\cite{Fernandes2018, Allamanis2017}.
We compare the methods at two levels -- for the entire dataset (similar to the analysis presented for RQ1) and at the level of individual projects.

\begin{figure}[h!]
    \centering
    \includegraphics[width=\linewidth]{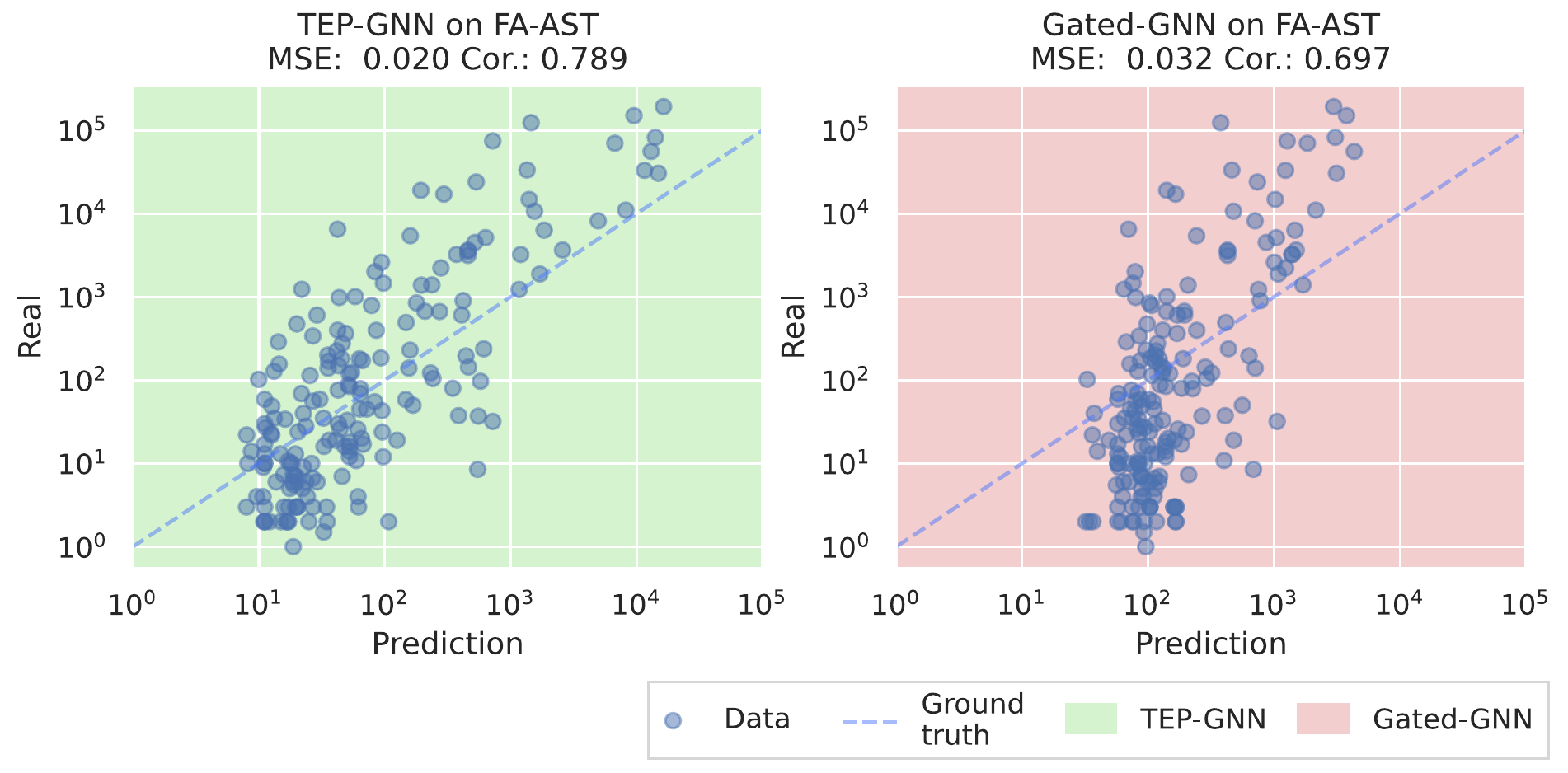}
    \caption{Comparison of \approach and a baseline (applying GGNN to the same FA-AST graphs). Dot points show real (y axes) and predicted (x axes) execution times produced by our model. The dash line refers to the perfect prediction.}
    \label{fig:astmodel}
\end{figure}

\paragraph{\textbf{Comparison for the entire dataset}}
We first apply both \approach and the baseline method to the dataset consisting of all projects. Figure \ref{fig:astmodel} depicts the respective results. Our model outperforms the baseline, with a Pearson correlation that is higher almost up to 0.1 (i.e., 0.789 versus 0.697). Hence, we conclude that our model and GNN architecture is indeed more appropriate to predict the execution time of test cases than a more standard GGNN approach.

Analyzing the results, we observe that \approach is able to achieve highly accurate predictions in most cases. However, there are rare outliers where our prediction model misses by approximately 20\%. The baseline GGNN method, on the other hand, has a tendency to predict fairly uniform execution times between $10^2$ and $10^3$, almost independent of what the actually observed test execution time is. Hence, it suffers from lower accuracy scores.

\begin{figure*}[h!]
    \centering
    \includegraphics[width=\textwidth]{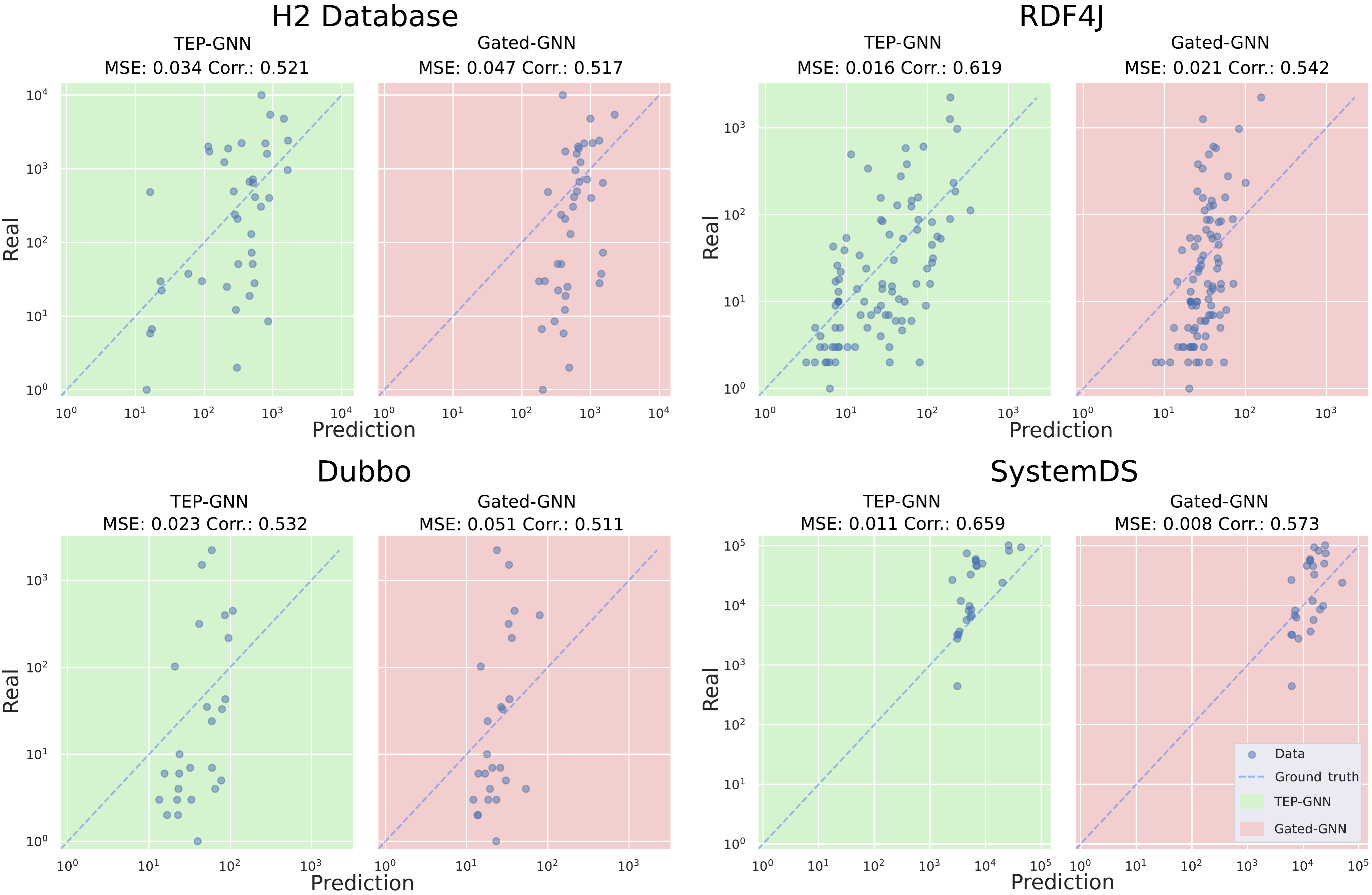}
\caption{Overview of \approach and the GGNN baseline trained for each individual project.}
\label{fig:projects}
\end{figure*}

\paragraph{\textbf{Comparison for individual projects}}
In the next step, we conduct a similar analysis, but focused on individual projects. This study answers the question of how well \approach works if trained on and used by a single project.
Thus, we train and test \approach and the baseline on each of the four projects individually. The results for each project are depicted in Figure~\ref{fig:projects}.

%\begin{figure*}[h!]
%\centering
%\begin{subfigure}{0.49\textwidth}
%    \includegraphics[width=\linewidth]{figures/single_datasetH2Database.pdf}
%    \caption{H2 database}
%    \label{fig:modelh2}
%\end{subfigure}
%\hfill %%
%\begin{subfigure}{0.49\textwidth}
%    \includegraphics[width=\linewidth]{figures/single_datasetRDF4J.pdf}
%    \caption{RDF4J}
%    \label{fig:modelrdf4j}
%\end{subfigure}
%\hfill %%
%\begin{subfigure}{0.49\textwidth}
%    \includegraphics[width=\linewidth]{figures/single_datasetDubbo.pdf}
%    \caption{Dubbo}
%    \label{fig:modeldubbo}
%\end{subfigure}
%\hfill %%
%\begin{subfigure}{0.49\textwidth}
%    \includegraphics[width=\linewidth]{figures/single_datasetSystemDS.pdf}
%    \caption{SystemDS}
%    \label{fig:modelsystemds}
%\end{subfigure}
%\caption{Overview of \approach and the GGNN baseline trained for each individual project.}
%\label{fig:projects}
%\end{figure*}

We observe that in general the prediction quality is substantially lower if the model is trained on individual projects, both for \approach and the baseline. \approach still outperforms the baseline for each project, but only with negligible prediction performance differences in the case of H2 and Dubbo. For RDF4J, which contains the largest number of test cases (and, consequently, the largest number of graphs to learn from), the difference between our approach and the baseline remains larger. 

From these results we conclude  that (a) \approach indeed outperforms the baseline in all the settings we tested, but (b) our approach works best if sufficient training graphs are available in comparison to the size of the graphs and vocabulary (if graphs are complex and/or training data is sparse the difference between our approach and the baseline is insignificant); (c) finally, we conclude that both approaches appear to learn some transferable knowledge even when training on graphs that originate from a different project.

% According to the observed results, the prediction quality is much lower than for the models trained and tested on the full dataset, which is implying that the model does indeed learn something that generalizes across projects. our \approach approach is still outperform the competitor on the level of the individual project. Figures \ref{fig:modelh2},\ref{fig:modeldubbo} clarify that for these smaller datasets like \textit{H2database, dubbo} the quality difference between our approach and the baseline is not exactly huge. Whereas and based on Figures \ref{fig:modelsystemds},\ref{fig:modelrdf4j} the \textit{RDF4J} which contains the highest test files. This means that our GNN model only becomes really useful once you have more data. 

%\todo[inline]{I guess the correlation score is also an important information to be added to Figures 8 and 9}
%\todo[inline]{\textbf{This is still missing:} Antonio the reviewer may ask why the real value and the predicted value are between 0,1. it is because the normalization that we did on the dependant valuable. please mentioned in Section 2.4 why you did the scaling of the execution time values. DONE}

\subsubsection{RQ3: Evaluation of Models Trained on Different Projects}
\begin{figure*}[t]
    \centering
    \includegraphics[width=\linewidth]{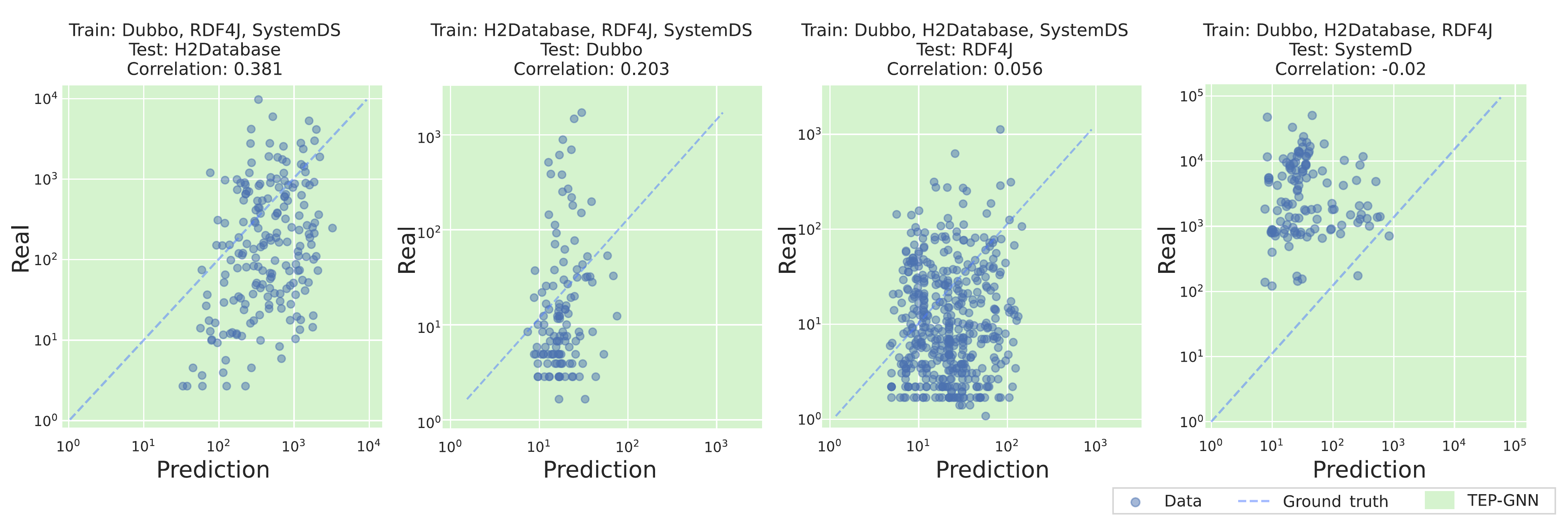}
    \caption{Testing trained models on unseen projects.}
    \label{fig:cross_projects}
\end{figure*}

%\todo[inline]{overview Table that contains all cross project analysis. MSE or Correlation score?! }
Based on the third conclusion above, we now raise the question if it is possible to build a generic model trained on a subset of projects and apply it to a new (unseen) project.
Hence, we now train new \approach models using three of the projects in our dataset, and test on the fourth one. The results for all four combinations are shown in Figure~\ref{fig:cross_projects}.

% We need to do a cross-project analysis to answer the third research question. Since our dataset projects are diverse in the sense that they belong to different categories, we would like to check whether our model is general enough to be applied to source code related to different open source projects. To accomplish this task, we need to train our model based on three projects, test it on the remaining project from different domains, and check if we could get good results. 
%Thus we will do a one - one cross projects analysis by training our model and testing it on one project.

We observe that the quality of predictions in this setting is in general rather disappointing, ranging from a Pearson correlation of 0.381 when testing on the H2 database to -0.02 when testing on systemDS. Interestingly, H2, which has the structurally most complex test cases in our study subjects, seems to perform better with a transferred model than the other projects, where test cases tend to be simpler. This is likely due to the test cases in projects such as systemDS mainly consisting of calls to the specific system-under-test, about which a transferred model clearly cannot learn any execution time properties.

Despite these results, we conclude that the models that achieve the highest prediction performance are trained with data across multiple projects, including ones with both complex and simpler test cases. A more general model, i.e., a model that has been trained using data of multiple projects (including the one that it is being applied to) is able to outperform a "pure" model that is trained and tested on a single, individual project.

 \section{Discussion}
\label{sec:discussion}

Our study results show that the accurate prediction of execution times of test
suites is possible. This gives developers an early indication of the time required
to run the cases in the build process, deciding in the process if techniques such
as test case selection are required.

\subsection{Lessons Learned}

\textbf{FA-ASTs are a promising approach to represent source code for performance prediction.} Unlike previous work~\cite{Narayanan2010, Meng2017, Zhou2019}, our goal in this study was to treat performance prediction as a regression rather than a classification (slow or fast) problem. Our results in Section~\ref{sec:results} indicate that using flow augmentation we are able to achieve good prediction quality.
% , as a pure AST representation is not sufficiently expressive to capture the dynamics that tend to impact software performance.
Furthermore, more information could be added to the FA-AST, such as program dependency graphs. We speculate that this approach is also promising to predict the performance of more complex, arbitrary code; however, more specific experiments in this direction need to be carried out as future research.

% \subsection{The Advantage of Augmentation}
% The studies that tried to predict the actual performance  converted the problem into a classification problem because, as they claim, it is difficult to predict the absolute value of performance. Finding one rich representation that holds all kinds of information from the source code was the key to the success of our approach. More information might also be added to the augmented version of code representation called flow graphs or program dependencies graphs. So merging all kinds of source code representation will be the key asset to having wealthy data to solve very complicated tasks like absolute performance prediction 

\textbf{GraphConv substantially outperforms the more common GGNN models in performance prediction as long as sufficient data is available.}
As discussed in Section~\ref{sec:results}, our GraphConv based GNN model substantially outperforms GGNN, which is a currently commonly used graph neural network model in software engineering research~\cite{Fernandes2018, Allamanis2017}. However, this is only true if sufficient data is available -- when training models for individual projects, we observed that, due to the limited amount of training data available in these cases, the performance difference between our GraphConv based model and the GGNN baseline was minimal. We conclude that, as long as sufficient data is available, GraphConv should also be investigated in other software engineering contexts that make use of GNNs.

% \subsection{The Advantage of Defending a Robust Model}
% Having a vibrant data representation is not the only key but also thinking about a robust model that can learn from this representation and extract different patterns that help the model be more general to solve any other related problem. We validate the efficiency of our model by comparing it with the most used GGNN in a similar problem. Thus, our model outperforms the baseline models to constitute a robust solution for our problem with the augmentation process. 

\textbf{\approach in its current form is not able to generalize to unseen projects.} While \approach can effectively predict the execution time of unseen test files, it can only do so if \emph{other test cases from the same project} have been used to build the model. This implies that much of what the model learns about test case performance during training is project-specific and does not generalize to other projects. This is not surprising, given that the core of test cases consist of invocations to the system-under-test, which will be different from project to project. However, our results also demonstrate that training a model on test cases  originating from a range of different projects leads to better predictions than training on a single project. This indicates that some cross-project learning indeed happens. Evidently, for practitioners a general model that can be applied to any project, without the need to train first based on historical executions from the same project, would be much more directly useful. Hence, future research should investigate whether approaches such as meta-learning~\cite{Joaquin2018} could be used to build more transferable models. At the very least, we hypothesize that general prediction models for project families, such as Apache or Eclipse projects, could be built. 

% \subsection{Cross-Project Dilemma}
% \todo[inline]{Morteza please give some ideas about meta learning and transfer learning by discussing other examples from other domains like BERT in NLP and DenseNet in computer vision. More info about Meta learning is also required}

% The results obtained by cross-project analysis are not promising due to the difference in the code patterns of the projects since the projects are related to a different domain. We claim that our solution is just proof of concepts that we can predict the performance of test cases of many projects. Thus, transfer learning and meta-learning might be used to build upon our solution to make our model more transferable to be successfully applied to many other projects and domains. On that basis, more data related to a different project is also needed to improve the quality of our model. Especially GNNs required a lot of samples to learn efficiently. 
% Moreover, despite the bad correlation results, we however infer a very important points:
% \begin{itemize}
%     \item The density of the code massively contribute in the prediction process. Its contribution is higher than the number of samples. 
%     \item one reason of the poor analysis is that  GNNs usually are data-hungry models.
%     \item projects that related to same apache and to the same domain but differnt application could lead to better results if we have rich model. 
% \end{itemize}

\subsection{Threats to Validity}
Similar to other experimental studies in software engineering, our work is subject to certain limitations and threats to validity, which we elaborate in the following.

\textbf{Internal Validity Threats.}
A key design choice in our study was the usage of existing, real-world data from GitHub's build system, rather than collecting performance data ourselves (e.g., on a dedicated experiment machine). This has obvious advantages with regards to the realism of our approach, but raises the threat that our training and test data may be subject to confounding factors outside of our knowledge. In particular, prior research has shown that even identically configured cloud virtual machines can vary significantly in performance~\cite{leitner:16}. However, the high accuracy achieved by our prediction models indicates that this is not a major concern with the data we used. That said, we expect that \approach would perform even better on the performance data that has been measured more rigorously.

Another design choice was that we predict execution times for entire test classes (files). More fine-grained predictions (e.g., for individual test cases in a class) would of course be doable, for instance by constructing the FA-AST with test methods as entry points rather than for an entire class as compilation unit. However, individual test cases often have very short execution times in relation to the precision with which build systems typically measure execution times (milliseconds), and the resulting graphs would be very small. 
We argue that our choice of test class granularity constitutes a good trade-off that is still useful for developers.

\textbf{External Validity Threats.}
An obvious question raised by our work is how well the results reported in Section~\ref{sec:results} would generalize to other projects. To mitigate this threat, we have chosen four relatively different Java projects as study subjects following a diversity sampling strategy~\cite{baltes:22}. However, our study does not allow us to conclude whether the \approach approach would generalize to other programming languages or closed-source software.

 \section{Related Work}
\label{sec:rw}
% We now discuss related work and how our research contributes to the body of research.

% \subsection{Performance Prediction with Deep Learning}
%\todo[inline]{TODO Peter and Philipp }
% Papers from intro
\textbf{Predicting software performance.}
Predicting the absolute value of a performance counter, such as execution time, based on the source code alone is challenging, as application performance is a function of several unknowns stemming from the application run-time and interactions between the OS and underlying hardware. This makes the problem notoriously challenging for any machine learning model, including deep learning techniques. Hence, existing studies often struggle with poor prediction accuracy~\cite{Meng2017,Narayanan2010}. One way to simplify the problem (and hence make it more tractable) is to convert it into a classification problem. Examples of this approach include Zhou et al.~\cite{zhou:19}, who predict if a program from a programming competition website exceeds the time limit, Ramadan et al.~\cite{Ramadan2021}, who predict whether a performance change is introduced by a code structure change, or Laaber et al.~\cite{laaber:21}, who have shown that a categorical classification of benchmarks into high- or low-variability is feasible.

 However, recent research has shown that predicting absolute performance values can be feasible in more specialized contexts. For example, Guo et al. successfully predict the execution time of a specific untested configuration of a configurable system~ \cite{guo:13,guo:18}, and Samoaa and Leitner have shown that the execution time of a benchmark with a specific workload configuration can be predicted~\cite{samoaa:21}. In this work, our core contribution is  we demonstrate that predicting absolute performance values is possible in another context, namely for the execution time of test files.

% Testing with Fewer Resources: An Adaptive Approach to Performance-Aware Test Case Generation
% https://ieeexplore.ieee.org/abstract/document/8865437
%\cite{grano:21}

% Ernest: Efficient Performance Prediction for Large-Scale Advanced Analytics
% https://www.usenix.org/conference/nsdi16/technical-sessions/presentation/venkataraman
% (not sure how related that and the next paper really are, but  heavily cited)
%\cite{Venkataraman:16,maros:19}

% % Machine Learning for Performance Prediction of Spark Cloud Applications
% % https://ieeexplore.ieee.org/abstract/document/8814514
% \cite{maros:19}

% % Comparative Code Structure Analysis using Deep Learning for Performance Prediction
% % https://www.computer.org/csdl/proceedings-article/ispass/2021/864300a151/1taFhsH8A36
% \cite{ramadan:21}

% The Interplay of Sampling and Machine Learning for Software Performance Prediction
% https://ieeexplore.ieee.org/abstract/document/9062326
%\cite{Kaltenecker:20}

% Does Configuration Encoding Matter in Learning Software Performance? An Empirical Study on Encoding Schemes
% https://arxiv.org/abs/2203.15988

% % Mira: A Framework for Static Performance Analysis
% % https://ieeexplore.ieee.org/document/8048922
% \cite{meng:17}

% \subsection{Graph Neural Networks for Software Engineering}
\textbf{Graph neural networks for software engineering.} Graph Neural Networks (GNNs) constitute an up-and-coming machine learning model in the context of software engineering research~\cite{samoaa2022}.
Graphs are mathematical structures used to model pairwise relations between objects. A graph can be used to model a wide number of different domains, ranging from biology \cite{huber2007graphs}, face-to-face human interactions  \cite{longa2022efficient,longa2022neighbourhood}, or digital contact tracing \cite{cencetti2021digital}. 
Li et al \cite{Li2015} use a GRU cell in gated graph neural networks (GGNNs) for updating the nodes' states. To evaluate their model they run the model on a basic program and try to detect null pointers. Instead of having the whole program as an input Li et al. \cite{Li2015} use the memory heap states of the program to the model.
Since the original work by Li et al. \cite{Li2015}, GGNNs have become a commonly used tool for applying GNNs in software engineering.
One challenge that needs to be solved before any GNN approach can be applied for code-based software engineering research is how to represent a program as graph. 

%One straightforward way is to use control flow graphs. Phan et al. \cite{Phan2017} use graph convolutional networks for defect detection on control flow graphs in C, based on compiled assembly code.
Phan et al. \cite{Phan2017} use graph convolutional networks (GCNs) based on compiled assembly code to detect defects on control flow graphs in C.
Another application of control flow graphs is using graph matching networks (GMN) between two graphs of binary functions proposed by Li et al. \cite{li19d}. 
%Li et al. \cite{li19d} proposed graph matching networks (GMN) for learning the similarity between two graphs. They applied their model to compute the similarity between control flow graphs of binary functions. 
Other researchers propose the creation of program graphs based on the AST. Allamanis et al. \cite{Allamanis2017} and Brockschmidt et al. \cite{Brockschmidt2018} use GGNN in C\# for naming variables and generating program expressions for code completion respectively.
%, similar to the FA-AST approach we are using in our work. 
%

%Allamanis et al. \cite{Allamanis2017} used GGNN to learn representations for C\# programs for two tasks: variable naming and correcting variable misuse. Brockschmidt et al. \cite{Brockschmidt2018} used GGNN to generate program expressions for code completion in C\#.
 \section{Conclusion and Future Work}
\label{sec:conclusions}

Inn this work we presented \approach, an effective method for predicting the execution time of Java test files. Our approach leverages explicitly capturing control and data flow information as augmentations to the program AST. Further, our approach applies high order convolution graph neural networks over this flow-augmented AST (FA-AST). By building FA-AST using original ASTs and flow edges, our approach can directly capture the syntax and semantic structure of test classes. Experimental results on four diverse test subjects demonstrate that by combining graph neural networks and control/data flow information, we can predict absolute test execution times with high accuracy.

%\todo[inline]{Morteza please check this paragraph and revise it. Morteza: I checked this and made some small changes.}
As the future work, we plan to further extent the FA-AST model currently used by \approach, as well as explore other ways of program representation to capture more syntactic and semantic code features.
Additionally, we plan to apply our approach to the execution time of general-purpose programs rather than test cases.
Finally, we would like to extend our current labeled data set by applying active learning to increase the amount of training data in a systematic way.

\section*{Acknowledgements}
This work received financial support from the Swedish Research Council VR under grant number 2018-04127 (Developer-Targeted Performance Engineering for Immersed Release and Software Engineering).

\bibliographystyle{abbrv}
\bibliography{references}

\end{document}